\documentclass[11pt]{article}
\usepackage{latexsym}
\usepackage{amsmath} % it subsumes: amstext, amsbsy, amsopn
\usepackage{amscd}     \usepackage{amsxtra}
\usepackage{upref}     \usepackage{amsthm}
\usepackage{amssymb}   %\usepackage{mathrsfs}
\usepackage{amsfonts}
\begin{document}

\title{{\bf Integrability-Nonintegrability Structures \\ and Individual
Photons' Description \\ as Finite Field Objects}}

\author{{\bf Stoil Donev}\footnote{e-mail:
 sdonev@inrne.bas.bg} \\ Institute for Nuclear Research and Nuclear
Energy,\\ Bulg.Acad.Sci., 1784 Sofia, blvd.Tzarigradsko chaussee 72,
Bulgaria\\
%\vskip 0.4cm
{\bf Maria Tashkova}\\Institute for Nuclear Research and Nuclear Energy,\\
Bulg.Acad.Sci., 1784 Sofia, blvd.Tzarigradsko chaussee 72, Bulgaria}
\date{}
\maketitle

\begin{abstract}
This paper presents an attempt to come to a natural field model of individual
photons considered as finite entities and propagating along some
distinguished direction in space in a consistent translational-rotational
manner. The starting assumption reflects their most trustful property to
propagate translationally in a uniform way along straight lines. The model
gives correct energy-momentum characteristics and connects the rotational
characteristics of photons with corresponding nonintegrability (or curvature)
of some 2-dimensional distributions (or Pfaff systems) on $\mathbb{R}^4$. It
is obtained that the curvature is proportional to the corresponding
energy-density. The field equations are obtained through a Lagrangian and
they express a consistency condition between photon's translational and
rotational propagation properties. The energy tensor is deduced directly from
the equations since the corresponding Hilbert energy-tensor becomes zero on
the solutions. Planck's formula $E=h\nu$ is naturally obtained as an
integral translational-rotational consistency relation.

\end{abstract}

PACS:

\section{Introduction}
At the very dawn of the 20th century M. Planck proposed [1] and a little bit
later Einstein appropriately used [2] the well known and widely used through
the whole last century simple formula $E=h\nu$, $h=const>0$. This formula
marked the beginning of a new era and became a real symbol of the physical
science during the following years. According to the Einstein's
interpretation it gives the full energy $E$ of {\it really existing} light
quanta of frequency $\nu$, and in this way a new understanding of the nature
of the electromagnetic field was introduced:  the field has structure which
contradicts the description given by Maxwell vacuum equations.  After De
Broglie's suggestion for the particle-wave nature of the electron obeying the
same energy-frequency relation [3], one could read Planck's formula in the
following way:  {\it there are physical objects in Nature the very existence
of which is strongly connected to some periodic (with time period $T=1/\nu$)
process of intrinsic for the object nature and such that the Lorentz
invariant product $ET$ is equal to $h$}. Such a reading should suggest that
these objects do NOT admit point-like approximation since the relativity
principle for free point particles requires straight-line uniform motion,
hence, no periodicity should be allowed. Instead, together with the notion
for {\it finiteness} of these objects and the hypothesis for admissible
point-like approximation all this led somehow to the dualistic and
probabilistic viewpoint of quantum mechanics, where these (free) objects,
except light quanta, have been considered as point-like but their (quantum)
states have been described by (spatially infinite) cosine plane wave
solutions to the potential-free Schr\"odinger equation [4], and the
measurable physical quantities have appeared as eigen values of
correspondingly defined operators satisfying definite commutation relations.

Although the great (from pragmatic point of view) achievements of such an
approach, known as {\it quantum theory}, the great challenge to build an
adequate description of individual representatives of these objects,
especially of light quanta called by Lewis {\it photons} [5], is still to be
appropriately met since  the efforts made in this direction, we have to
admit, still have not brought satisfactory results. Recall that
Einstein in his late years recognizes [6] that "the whole fifty years of
conscious brooding have not brought me nearer to the answer to the question
"what are light quanta", and now, half a century later, theoretical
physics, although some existing attempts [7-19], still needs progress to
present a satisfactory answer to the question "what is a photon". We consider
the corresponding theoretically directed efforts as necessary and even {\it
urgent} in view of the growing amount of definite experimantal skills in
manipulation with individual photons, in particular, in connection with the
experimental advancement in the "quantum computer" project.  The dominating
modern theoretical view on microobjects is based on the notions and concepts
of quantum field theory (QFT) where the structure of the photon (as well as
of any other microobject) is accounted for mainly through the so called {\it
structural function}, and highly expensive and delicate collision experiments
are planned and carried out namely in the frame of these concepts and methods
(see the 'PHOTON' Conferences Proceedings, some recent review papers:
[20-23]). Going not in details we just note a special feature of this QFT
approach: if the study of a microobject leads to conclusion that it has
structure, i.e. it is not point-like, then the corresponding constituents of
this structure are considered as point-like, so the point-likeness stays in
the theory just in a lower level.

In this paper we follow another approach based on the assumption that the
description of the available (most probaly NOT arbitrary) spatial structure
of photons can be made by {\it continuous finite/localized} functions of the
three space variables.  The difficulties met in this approach consist mainly,
in our view, in finding adequate enough mathematical objects and solving
appropriate PDE.  The lack of sufficiently reliable corresponding information
made us look into the problem from as general as possible point of view on
the basis of those properties of photons which may be considered as most
undoubtedly trustful. The analysis made suggested that such a property seems
to be the fact that {\it the propagation of an individual photon necessarily
includes a straight-line uniform component}, so we shall focus on this
property in order to see what useful for our purpose suggestions could be
deduced and what appropriate structures could be constructed. All these
suggestions and structures should be the building material for a
step-by-step creating a {\it self-consistent} system, i.e. the corresponding
properties must live in harmony with each other. From theoretical viewpoint,
these properties should reflect various important aspects of some general
notion about individual photons, and we now turn to outline briefly such a
notion.

\section{The Notion}
We introduce the following concept of individual photons, and we consider
it as working one, so it may be subject to future improvements.

\begin{center}
{\bf Photons are real massless time-stable physical objects
with a consistent translational-rotational dynamical structure}.
\end{center}

We give now some explanatory comments, beginning with the term {\it real}.
{\bf First} we emphasize that this term means that we consider photons as
{\it really} existing {\it physical} objects, not as appropriate and helpful
but imaginary (theoretical) entities.  Accordingly, photons {\bf necessarily
carry energy-momentum}, otherwise, they could hardly be detectable.  {\bf
Second}, photons can undoubtedly be {\it created} and {\it destroyed}, so, no
point-like and infinite models are reasonable: point-like objects are assumed
to have no structure, so they can not be destroyed since there is no
available structure to be destroyed; creation of infinite objects (e.g.
plane waves) requires infinite quantity of energy to be transformed from one
kind to another for finite time-period, which seems also unreasonable.
Accordingly, photons are {\it spatially finite} and have to be modeled like
such ones, which is the only possibility to be consistent with their
"created-destroyed" nature. It seems hardly believable that spatially
infinite and indestructible physical objects may exist at all.  {\bf Third},
"spatially finite" implies that photons may carry only {\it finite values} of
physical (conservative or non-conservative) quantities.  In particular, the
most universal physical quantity seems to be the energy-momentum, so the
model must allow finite integral values of energy-momentum to be carried by
the corresponding solutions. {\bf Fourth}, "spatially finite" means also that
photons {\it propagate}, i.e.  they do not "move" like classical particles
along trajectories, therefore, partial differential equations should be used
to describe their evolution in time.

The term "{\bf massless}" characterizes the way of propagation: the
integral energy $E$ and momentum $p$ of a photon should satisfy the relation
$E=cp$, where $c$ is the speed of light in vacuum, and in relativistic terms
this means that their integral energy-momentum vector {\it must be
isotropic}, i.e. it must have zero module with respect to Lorentz-Minkowski
(psuedo)metric in $\mathbb{R}^4$. Since the translational velocity of every
point where the corresponding field functions are different from zero must
be equal to $c$, we have in fact an isotropic vector field
$\bar{\zeta}$. The integral trajectories of this vector field are isotropic
straight lines. It follows that just the corresponding direction is
important, so, canonical coordinates $(x^1,x^2,x^3,x^4)=(x,y,z,\xi=ct)$ on
$\mathbb{R}^4$ may be chosen such that $\bar{\zeta}$ may have only two
non-zero components of magnitude $1$:  $\bar{\zeta}^\mu=(0,0,-\varepsilon,
1)$, where $\varepsilon=\pm 1$ accounts for the two directions along the
cooedinate $z$. Further such a coordinate system will be called
$\bar{\zeta}$-adapted and will be of main usage. It may be also expectable,
that the corresponding energy-momentum tensor $T_{\mu\nu}$ of the model
satisfies the relation $T_{\mu\nu}T^{\mu\nu}=0$, which may be considered as a
localization of the integral isotropy condition $E^2-c^2p^2=0$.

The term "{\bf translational-rotational}" means that besides translational
component along $\bar{\zeta}$, the propagation demonstrates also some
rotational (in the general sense of this concept) component in such a way
that {\it both components exist simultaneously and consistently}. It seems
reasonable to expect that such kind of behavior should be consistent only
with some distinguished spatial shapes. Moreover, if the Planck relation
$E=h\nu$ must be respected throughout the evolution, the rotational component
of propagation should have {\it periodical} nature and one of the two
possible, {\it left} or {\it right}, orientations. It seems reasonable also
to expect periodicity in the spatial shape of photons, which somehow to be
related to the time periodicity.

The term "{\bf dynamical structure}" means that the propagation is supposed to
be necessarily accompanied with {\it internal energy-momentum
redistribution}, which may be considered in the model as energy-momentum
exchange between (or among) somewhat functionally isolated subsystems.  It
could also mean that photons live in a dynamical harmony with the outside
world, i.e.  {\it any outside directed energy-momentum flow should be
accompanied by a parallel inside directed energy-momentum flow}. From
theoretical point of view this would mean that the corresponding mathematical
model-object should be a many-component one, and most probably, the
components should be vectors, tensors, etc. In other words, the mathematical
model-object may have two or more independent systems of indices.

Finally, note that if the time periodicity and the spatial periodicity should
be consistent somehow, the simplest such consistency would seem like this:
the spatial longitudinal dimension $2\pi l_o$ is equal to $cT$: $2\pi
l_o=cT$, where $l_o$ is some finite positive characteristic constant of the
corresponding solution. This would mean that every individual photon
determines its own length/time scale.

From differential viewpoint the translational-rotational consistency
should mean that every translational change (i.e. change along $\bar{\zeta}$)
of the mathematical object $F$ representing an individual photon, which change
is naturally given by the Lie derivative $L_{\bar{\zeta}}F$ of $F$ along
$\bar{\zeta}$, must be accompanied by a "rotational" change $D(F)$ of $F$,
where $D$ is some rotation generating operator. So, we should expect the
appearance of a relation stating some simple and strong dependence (e.g.
proportionality) between $\kappa L_{\bar{\zeta}}F$ and  $D(F)$, where
$\kappa=\pm1$ accounts for the {\it left/right} rotational orientations.

According to the above $\bar{\zeta}$ turns out to be the most natural
and trustful mathematical object that has local nature and that could be used
to induce other appropriately consistent and physically useful local
geometric structures on $\mathbb{R}^4$.  So, in the next section we are going
to study some of these structures.

\section{Some Geometry on $\mathbb{R}^4$ induced by $\bar{\zeta}$}

We consider the space $\mathbb{R}^4$ as a manifold related to standard global
coordinates
$$
(x^1,x^2,x^3,x^4)=(x,y,z,\xi=ct)
$$
and shall not make use of the
Minkowski metric for the time being. The only structures on
$\mathbb{R}^4$ assumed to exist for now are the vector field $\bar{\zeta}$
and the natural volume form $\omega_o=dx\wedge dy\wedge dz\wedge d\xi$.
We choose the coordinates in such a way that $\bar{\zeta}$ looks as follows:
\begin{equation}
\bar{\zeta}=-\varepsilon\frac{\partial}{\partial z} +          %1%
\frac{\partial}{\partial \xi}, \ \ \varepsilon=\pm 1.
\end{equation}

Since every vector field on a manifold generates 1-dimensional
locally integrable differential system the same should be true for our field
$\bar{\zeta}$. Let's denote the corresponding 3-dimensional Pfaff system by
$\Delta^*(\bar{\zeta})$. Thus, $\Delta^*(\bar{\zeta})$ is generated by three
linearly independent 1-forms $(\alpha_1,\alpha_2,\alpha_3)$ which annihilate
$\bar{\zeta}$, i.e.
$$
\alpha_1(\bar{\zeta})=\alpha_2(\bar{\zeta})=\alpha_3(\bar{\zeta})=0; \ \
\alpha_1\wedge \alpha_2\wedge \alpha_3\neq 0.
$$
The following basis of $\Delta^*(\bar{\zeta})$ will be of use, and instead of
$(\alpha_1, \alpha_2, \alpha_3)$ we introduce the notation $(A, A^*, \zeta)$:
\begin{equation}
 A =udx+pdy; \ \ A^*=-pdx+udy; \ \ \zeta=\varepsilon dz+d\xi,      %2%
\end{equation}
where $(u,p)$ are two arbitrary functions.

{\bf Proposition 1}. The following relations hold:
\begin{equation}
\mathbf{d}A\wedge A\wedge A^*=0;\ \
\mathbf{d}A^*\wedge A^*\wedge A=0;\ \                              %3%
\end{equation}
\begin{equation}
\mathbf{d}A\wedge A\wedge \zeta=
\varepsilon\big[u(p_\xi-\varepsilon p_z)-
p(u_\xi-\varepsilon u_z)\big]\omega_o; \ \
\mathbf{d}A^*\wedge A^*\wedge \zeta=                                  %4%
\varepsilon\big[u(p_\xi-\varepsilon p_z)-
p(u_\xi-\varepsilon u_z)\big]\omega_o.
\end{equation}
These relations (3)-(4) say that the 2-dimensional Pfaff system $(A,A^*)$ is
completely integrable for any choice of the two functions $(u,p)$, while the
two 2-dimensional Pfaff systems $(A,\zeta)$ and $(A^*,\zeta)$ are NOT
completely integrable in general, and the same curvature factor \linebreak
$\mathbf{R}=u(p_\xi-\varepsilon p_z)-p(u_\xi-\varepsilon u_z)$
determines their nonintegrability.

In order to give the dual non-integrability picture we consider
the 1-form $\zeta$. It determines a 3-dimensional distribution (or
differential system) $\Delta(\zeta)$, and a basis of this distribution is
given by the vector fields $(\bar{A},\bar{A^*},\bar{\zeta})$, where
\begin{equation}
\bar{A}=-u\frac{\partial}{\partial x}-p\frac{\partial}{\partial y};\ \
\bar{A^*}=p\frac{\partial}{\partial x}-u\frac{\partial}{\partial y}; \ \
\end{equation}
and the vector field $\bar{\zeta}$ is given by (1): \ \ $\zeta(\bar{A})=
\zeta(\bar{A^*})=\zeta(\bar{\zeta})=0; \
\bar{A}\wedge \bar{A^*}\wedge\bar\zeta\neq 0$.

Further in the paper we shall relate the 2-forms on $\mathbb{R}^4$ to the
basis
\[
dx\wedge dy,\ \ dx\wedge dz,\ \ dy\wedge dz,\ \
dx\wedge d\xi,\ \ dy\wedge d\xi,\ \ dz\wedge d\xi.
\]
It deserves to note here that one of the canonical complex structures in the
bundle of 2-forms, denoted by $\mathcal{J}$, and given by
\[
\mathcal{J}(dx\wedge dy)=-dz\wedge d\xi, \ \
\mathcal{J}(dx\wedge dz)=dy\wedge d\xi, \ \
\mathcal{J}(dy\wedge dz)=-dx\wedge d\xi,
\]
\[
\mathcal{J}(dx\wedge d\xi)=dy\wedge dz, \ \
\mathcal{J}(dy\wedge d\xi)=-dx\wedge dz, \
\ \mathcal{J}(dz\wedge d\xi)=dx\wedge dy
\]
transforms $A\wedge\zeta$ to $A^*\wedge\zeta$.

{\bf Proposition 2}. The following Lie-bracket relations hold:
\begin{equation}
[\bar{A},\bar{\zeta}]=(u_\xi-\varepsilon u_z)\frac{\partial}{\partial x}+
(p_\xi-\varepsilon p_z)\frac{\partial}{\partial y};            %6%
\end{equation}
\begin{equation}
[\bar{A^*},\bar{\zeta}]=-(p_\xi-\varepsilon p_z)\frac{\partial}{\partial x}+
(u_\xi-\varepsilon u_z)\frac{\partial}{\partial y}.            %7%
\end{equation}
In accordance with {\bf Prop.1} from these last relations (6)-(7) it follows
that the two distributions $(\bar{A},\bar{\zeta})$ and
$(\bar{A^*},\bar{\zeta})$ would be completely integrable only if the same
curvature factor
\begin{equation}
\mathbf{R}=u(p_\xi-\varepsilon p_z)-p(u_\xi-\varepsilon u_z)     %8%
\end{equation}
is zero. In fact, for example, if $(\bar{A},\bar{\zeta})$ is completely
integrable then there must exist two functions $f$ and $g$ such that the Lie
bracket $[\bar{A},\bar{\zeta}]$ must be representable as follows:
$$
[\bar{A},\bar{\zeta}]=f\bar{A} + g\bar{\zeta}.
$$
Since $[\bar{A},\bar{\zeta}]$ has no components along
$\frac{\partial}{\partial z}$ and $\frac{\partial}{\partial \xi}$ the above
relation reduces to
$$
(u_\xi-\varepsilon u_z)\frac{\partial}{\partial x}+
(p_\xi-\varepsilon p_z)\frac{\partial}{\partial y}=
f\bar{A}=
-fu\frac{\partial}{\partial x}-fp\frac{\partial}{\partial y}.
$$
From this last relation it follows
$$
-f=\frac{u_\xi-\varepsilon u_z}{u}=\frac{p_\xi-\varepsilon p_z}{p},
$$
i.e. $u(p_\xi-\varepsilon p_z)-p(u_\xi-\varepsilon u_z)=\mathbf{R}=0$.

We mention also that the projections
$$
\langle A,[\bar{A^*},\bar{\zeta}]\rangle=
-\langle A^*,[\bar{A},\bar{\zeta}]\rangle=
u(p_\xi-\varepsilon p_z)-p(u_\xi-\varepsilon u_z)
$$
give the same factor $\mathbf{R}$. The same curvature factor
appears, of course, as coefficient in the exterior products
$$
[\bar{A^*},\bar{\zeta}]\wedge \bar{A^*}\wedge\bar{\zeta} \ \ \ \text{and} \ \
\ [\bar{A},\bar{\zeta}]\wedge \bar{A}\wedge\bar{\zeta}.
$$
For the other two projections we obtain
\begin{equation}
\langle
A,[\bar{A},\bar{\zeta}]\rangle=                          %9%
-\langle
A^*,[\bar{A^*},\bar{\zeta}]\rangle= \frac12\big[(u^2+p^2)_\xi-\varepsilon
(u^2+p^2)_z\big].
\end{equation}
Clearly, the last relation (9) may be put in
terms of the Lie-derivative as
$$
\frac 12L_{\bar{\zeta}}(u^2+p^2)=
-\frac12L_{\bar{\zeta}}\langle A,\bar{A}\rangle=
-\langle A,L_{\bar{\zeta}}\bar{A}\rangle=
-\langle A^*,L_{\bar{\zeta}}\bar{A^*}\rangle.
$$

{\bf Remark}. Further in the paper we shall denote
$\sqrt{u^2+p^2}\equiv \phi$, and shall assume that $\phi$ is a {\it spatially
finite} function, so, $u$ and $p$ must also be spatially finite.

It deserves to note that the tensor field $T=\phi^2 \zeta\otimes\bar{\zeta}$,
considered as a linear map, satisfies the relation $T\circ T=0$, so it is a
boundary operator with image space generated by $\bar{\zeta}$ and kernel
space generated by the vectors $(\bar {A},\bar {A^*}, \bar{\zeta})$, so, the
corresponding homology space is generated by the classes of $\bar{A}$ and
$\bar{A^*}$. Moreover, in this coordinate system we have $T^{4}_{4}=\phi^2$
and $tr(T\circ T)=T^\nu_\mu T_\nu^\mu=0$, so $T$ seems to be an appropriate
candidate for energy-momentum tensor of a photon-like solution. As it is
easily seen a pseudoeuclidean metric $\eta$ in the tangent bundle generates a
euclidean metric $g$ on this homology space such that on the classes
$[\bar{A}]$ and $[\bar{A^*}]$ we have $|[\bar{A}]|=
\sqrt{|\eta(\bar{A},\bar{A})|}$, and
$|[\bar{A^*}]|=\sqrt{|\eta(\bar{A^*},\bar{A^*})|}$.

{\bf Proposition 3.} There is a function $\psi(u,p)$ such, that

$$L_{\bar{\zeta}}\psi=\frac
{u(p_\xi-\varepsilon p_z)-p(u_\xi-\varepsilon u_z)}{\phi^2}=
\frac{\mathbf{R}}{\phi^2} .
$$
\vskip 0.4cm
{\bf Proof}. It is immediately checked that $\psi=\arctan\frac pu$
is such one.
\vskip 0.4cm
We note that the function $\psi$ has a natural
interpretation of {\it phase} because of the easily verified now relations
$$
u=\phi\cos\psi,\ \ \ p=\phi \sin\psi,
$$
and $\phi$ acquires the status of {\it amplitude}. Since the transformation
$(u,p)\rightarrow (\phi,\psi)$ is non-degenerate this allows to
work with the two functions $(\phi,\psi)$ instead of $(u,p)$.

Note that the amplitude $\phi$ appears as square root of the determinant
function of the basis transformation
$$
\left\{\frac{\partial}{\partial x}, \frac{\partial}{\partial y},
\frac{\partial}{\partial z}, \frac{\partial}{\partial \xi}\right\}\rightarrow
\left\{\bar{A},\bar{A^*},\frac{\partial}{\partial z},
\frac{\partial}{\partial \xi}\right\},
$$
and the function $\psi$ appears as $arc\cos(1-\frac12\theta)$
where $\theta$ is the trace of the basis transformation
$$
\left\{\frac{\partial}{\partial x},
\frac{\partial}{\partial y}, \frac{\partial}{\partial z},
\frac{\partial}{\partial \xi}\right\}\rightarrow
\left\{\frac{\bar{A}}{\phi},
\frac{\bar{A^*}}{\phi},\frac{\partial}{\partial z},
\frac{\partial}{\partial \xi}\right\}.
$$
It follows that $\phi$ and $\psi$ have well defined invariant sense.

Finally we note that the same curvature factor may be obtained through
declaring the subspaces generated by $(\bar{A},\bar{\zeta})$ and
$(\bar{A^*},\bar{\zeta})$ as {\it horizontal}, and the subspaces generated
respectively by $\left(\bar{A^*},\varepsilon \frac{\partial}{\partial z}+
\frac{\partial}{\partial \xi}\right)$ and $\left(\bar{A},\varepsilon
\frac{\partial}{\partial z}+ \frac{\partial}{\partial \xi}\right)$ as {\it
vertical}, and computing the corresponding vertical projections,
respectively, of $[\bar{A},\bar{\zeta}]$ and  $[\bar{A^*},\bar{\zeta}]$.

From {\bf Prop.3} we have
\begin{equation}
\mathbf{R}=\phi^2L_{\bar{\zeta}}\psi= \
\phi^2(\psi_\xi-\varepsilon\psi_z) \,.             %10%
\end{equation}

This last formula (10) shows something very important: at any $\phi\ne 0$ the
curvature $\mathbf{R}$ will NOT be zero only if the phase $\psi$ is NOT a
running wave along $\bar{\zeta}$, i.e. only if $L_{\bar{\zeta}}\psi\neq 0$,
which admits in principle availability of rotation. In fact, lack of rotation
would mean that $\phi$ and $\psi$ are running waves along $\bar{\zeta}$.
The relation $L_{\bar{\zeta}}\psi\neq 0$ means, however, that rotational
properties are possible in general, and some of these properties are
carried by the phase $\psi$. It follows that in such a case the
translational component of propagation along $\bar{\zeta}$ (which is supposed
to be available) must be determined essentially, and most probably entirely,
by $\phi$.  In particular, we could expect the relation
$L_{\bar{\zeta}}\phi=0$ to hold, and if this happens, then the rotational
component of propagation will be represented entirely by the phase $\psi$,
and, more specially, by the curvature factor $\mathbf{R}\neq 0$, so, further
we assume that, in general, $L_{\bar{\zeta}}\psi\neq 0$.

We consider now the Lie-brackets $[\bar{A},\bar{\zeta}]$ and
$[\bar{A^*},\bar{\zeta}]$ in terms of $\psi$:
\[
[\bar{A},\bar{\zeta}]=-L_{\bar{\zeta}}\psi.\bar{A^*} \ ; \ \
[\bar{A^*},\bar{\zeta}]=L_{\bar{\zeta}}\psi.\bar{A} \  ; \ \
[\bar{A},\bar{\zeta}]\wedge[\bar{A^*},\bar{\zeta}]=
-\left(L_{\bar\zeta}\psi\right)^2\bar{A^*}\wedge\bar{A}\neq 0 .
\]
These relations say that the 2-dimensional frame $(\bar{A},\bar{A^*})$ on
the $(x,y)$-plane is transformed to the 2-dimensional
frame $([\bar{A},\bar{\zeta}],[\bar{A^*},\bar{\zeta}])$ on the same plane by
means of the matrix $L_{\bar{\zeta}}\psi.J$, where $J$ is the canonical
complex structure in $\mathbb{R}^2$, generating rotation to the angle of
$\pi/2$.  Hence, the 4-frame $(\bar{A},\bar{A^*},\partial_z,\partial_\xi)$ is
transformed to the 4-frame
$([\bar{A},\bar{\zeta}],[\bar{A^*},\bar{\zeta}],\partial_z,\partial_\xi)$ by
means of a linear map with determinant $(L_{\bar\zeta}\psi)^2$. This
observation confirms once again that at $\phi\neq 0$ {\bf the rotational
component of propagation will be available only if
$L_{\bar{\zeta}}\psi\neq 0$, i.e.  only if the non-integrability factor
$\mathbf{R}$ is non-zero}.  Moreover, it is suggested to make use of some
complex structures as possible rotation-generating operators.

The two nonintegrable Pfaff systems $(A,\zeta)$ and $(A^*,\zeta)$ define
corresponding 2-forms: \linebreak
$G=A\wedge\zeta$ and $G^*=A^*\wedge\zeta$.
Let's denote $\bar{A}\wedge\bar{\zeta}\equiv \bar{G}$,
$\bar{A^*}\wedge\bar{\zeta}\equiv\bar{G^*}$.
We shall need the quantities
$i(\bar{G})\mathbf{d}G+i(\bar{G^*})\mathbf{d}G^*$ and
$i(\bar{G^*})\mathbf{d}G+i(\bar{G})\mathbf{d}G^*$
, where
$i(\bar{G})=i(\bar{\zeta})\circ i(\bar{A})$, and, analogically,
$i(\bar{G^*})=i(\bar{\zeta})\circ i(\bar{A^*})$,
and $i(X)$ is the standard insertion operator in the exterior algebra
of differential forms on $\mathbb{R}^4$ defined by the vector field $X$.

Having in view the explicit expressions for $A,A^*,\zeta,\bar{A},\bar{A^*}$
and $\bar{\zeta}$, and making use of the coordinate free formula for the
exterior derivative of 1-form $\alpha$:
$$
\mathbf{d}\alpha(X,Y)=X\langle\alpha,Y\rangle-
Y\langle\alpha,X\rangle-\alpha([X,Y])
$$
we obtain
\[
i(\bar{G})\mathbf{d}G=i(\bar{G}^*)\mathbf{d}G^*=
\frac12 L_{\bar{\zeta}}\left(\phi^2\right).\,\zeta \  ,
\]
\begin{equation}
i(\bar{G})\mathbf{d}G+i(\bar{G^*})\mathbf{d}G^*
=\frac12\Big[(\bar{G})^{\alpha\beta}(\mathbf{d}G)_{\alpha\beta\mu}     %11%
+(\bar{G^*})^{\alpha\beta}(\mathbf{d}G^*)_{\alpha\beta\mu}\Big]dx^\mu
=L_{\bar{\zeta}}\left(\phi^2\right).\,\zeta \ .\ \
\end{equation}
Also, we have
\[
i(\bar{G^*})\mathbf{d}G=-i(\bar{G})\mathbf{d}G^* \ ,
\]
\begin{equation}
i(\bar{G^*})\mathbf{d}G+i(\bar{G})\mathbf{d}G^*=                       %12%
\frac12\Big[(\bar{G^*})^{\alpha\beta}(\mathbf{d}G)_{\alpha\beta\mu}
+(\bar{G})^{\alpha\beta}(\mathbf{d}G^*)_{\alpha\beta\mu}\Big]dx^\mu=0 \ .
\end{equation}
A direct calculation shows that
\[
i(\bar{G^*})\mathbf{d}G=-i(\bar{G})\mathbf{d}G^*=
\]
\[
=\varepsilon\Big[u(p_\xi-\varepsilon p_z)-p(u_\xi-\varepsilon u_z)\Big]dz+
\Big[u(p_\xi-\varepsilon p_z)-p(u_\xi-\varepsilon u_z)\Big]d\xi
=\mathbf{R}.\,\zeta \ .
\]
We note that if $\eta$ is the pseudoeuclidean metric with signature
$(-,-,-,+)$, then making use of the corresponding Hodge $*$-operator, the
following relations can be easily verified:
\[
(\varepsilon G^*)_{\mu\nu}=
(*G)_{\mu\nu}, \ \bar{A}^\mu=\eta^{\mu\nu}A_\nu, \
\bar{\zeta}^\mu=\eta^{\mu\nu}\zeta_\nu, \ \
\bar{G}^{\mu\nu}=\eta^{\mu\sigma}\eta^{\nu\tau}G_{\sigma\tau}, \ \
(\bar{\varepsilon G^*})^{\mu\nu}=
\eta^{\mu\sigma}\eta^{\nu\tau}(*G)_{\sigma\tau} \ .
\]
Moreover, with respect to the corresponding Levi-Civita covariant derivative
$\nabla$ we obtain
\begin{equation}
\nabla_{\nu}\Big[\frac14 G_{\alpha\beta}\bar{G}^{\alpha\beta}\delta^\nu_\mu-
G_{\mu\sigma}\bar{G}^{\nu\sigma}\Big]=
\frac12 \Big[(\bar{G})^{\alpha\beta}(\mathbf{d}G)_{\alpha\beta\mu}      %13%
+(\bar{*G})^{\alpha\beta}(\mathbf{d}*G)_{\alpha\beta\mu}\Big] \ .
\end{equation}
Relations (11) and (13) show that if $L_{\bar{\zeta}}\phi=0$, then the well
known Maxwell energy-momentum tensor, determined by the
2-form $G=A\wedge\zeta$ has zero divergence, and additionally, we get
\begin{equation}
\phi^2\zeta_\mu\bar{\zeta}^\nu= \frac14
G_{\alpha\beta}\bar{G}^{\alpha\beta}\delta^\nu_\mu-
G_{\mu\sigma}\bar{G}^{\nu\sigma}=
\frac12\Big[G_{\mu\sigma}\bar{G}^{\nu\sigma}+                      %14%
(*G)_{\mu\sigma}\bar{(*G)}^{\nu\sigma}\Big] \ .
\end{equation}

We consider now a consequence of having a completely integrable
1-dimensional Pfaff system on $\mathbb{R}^4$ defined by the 1-form $\omega :
\mathbf{d}\omega\wedge\omega=0$. This 1-dimensional Pfaff system is
determined up to a nonvanishing function $f$:  $\omega\rightarrow f\omega,
f(x)\neq 0, x\in M$, since $f\omega$ satisfies the same equation
$\mathbf{d}(f\omega)\wedge(f\omega)=0$. It follows
that there is 1-form $\theta$ (in fact a class of 1-forms
$(\theta+g\omega)$, $g$ is a function) such, that
$\mathbf{d}\omega=\theta\wedge\omega$.  Now, the Godbillon-Vey theorem
says that the 3-form $\beta=\mathbf{d}\theta\wedge\theta$ is closed:
$\mathbf{d}\beta=\mathbf{d}(\mathbf{d}\theta\wedge\theta)=0$, and, varying
$\theta$ and $\omega$ in an admissible way: $\theta\rightarrow
(\theta+g\omega); \ \omega\rightarrow f\omega$, leads to adding an
exact 3-form to $\beta$, so we have a cohomological class $\Gamma$ defined
entirely by the integrable 1-dimensional Pfaff system. From physical point
of view this seems to be important because it shows that each completely
integrable 1-dimensional Pfaff system on Minkowski space may generate a
conservation law through integrating over $\mathbb{R}^3$ the restriction of
$\beta $ to $\mathbb{R}^3$, provided this integral is finite.

Recall now our objects: $A$, $A^*$,
$\zeta$ and the corresponding vector fields
 $\bar{A}, \bar{A^*}, \bar{\zeta}$.
Let's consider the 1-form $\omega=f\zeta=\varepsilon fdz+fd\xi$, where $f$ is
a nonvanishing function on $M$. We have the relations:
\[
\omega(\bar{A})=0,\ \ \ \omega(\bar{A^*})=0,\ \ \ \omega(\bar{\zeta})=0.
\]
Moreover, since $\zeta$ is closed, $\omega=f\zeta$ satisfies the Frobenius
integrability condition:
\[
\mathbf{d}\omega\wedge\omega=f\mathbf{d}f\wedge\zeta\wedge\zeta=0.
\]
Therefore, the corresponding 1-dimensional Pfaff system, defined by
$\omega=f\zeta$, is completely integrable, and there exists a new 1-form
$\theta$, such that $\mathbf{d}\omega=\theta\wedge\omega$, and
$\mathbf{d}(\mathbf{d}\theta\wedge\theta)=0$.

Clearly, the natural imbedding $i: (x,y,z)\rightarrow(x,y,z,0)$ will give the
natural restriction $i^*\beta$ to $\mathbb{R}^3$. We note that from our
point of view it is not so important whether $\Gamma$ is trivial or
nontrivial. The important point is the 3-form $\beta$ to have appropriate
component $\beta_{123}$ in front of the basis element $dx\wedge dy\wedge dz$,
because only this component survives after the restriction considered is
performed, which formally means that we put $\xi=0$ and $d\xi=0$ in $\beta$.
The value of the corresponding conservative quantity will be found provided
the integration can be carried out successfully, i.e. when $(i^*\beta)_{123}$
is concentrated in a finite 3d-subregion of $\mathbb{R}^3$ with no
 singularities.

In order to find appropriate $\theta$ in our case we are going to take
advantage of the freedom we have when choosing $\theta$: the 1-form $\theta$
is defined up to adding to it an 1-form $\alpha=g\,\omega$, where $g$
is an arbitrary function on $M$, because $\theta$ is defined by the relation
$\mathbf{d}\omega=\theta\wedge\omega$, and
$(\theta+g\,\omega)\wedge\omega=\theta\wedge\omega$
always. The freedom in choosing $\omega$ consists in choosing the
function $f$, and we shall show that $f$ may be chosen in such a way:
$\omega=f\,\zeta$, that the corresponding integral of $i^*\beta$ to present a
finite conservative quantity.

Recalling that $\mathbf{d}\zeta=0$, we have
\[
\mathbf{d}\omega=\mathbf{d}(f\zeta)=\mathbf{d}f\wedge\zeta+f\mathbf{d}\zeta=
\mathbf{d}f\wedge\zeta.
\]
Since $\mathbf{d}\omega$ must be equal to $\theta\wedge\omega$ we obtain
\[
\mathbf{d}\omega=\mathbf{d}f\wedge\zeta=\theta\wedge\omega=
\theta\wedge(f\zeta)=f\theta\wedge\zeta.
\]
It follows
\[
\theta\wedge\zeta=
\frac{1}{f}\mathbf{d}f\wedge\zeta=\mathbf{d}(ln\,f)\wedge\zeta=
\Big[\mathbf{d}(ln\,f)+h\zeta\Big]\wedge\zeta,
\]
where $h$ is an arbitrary function. Hence, in general, we obtain
$\theta=\mathbf{d}(ln\,f)+h\zeta$. Therefore,
$\mathbf{d}\theta=\mathbf{d}h\wedge\zeta$ and for
$\beta=\mathbf{d}\theta\wedge\theta$ we obtain
\[
\beta=\mathbf{d}\theta\wedge\theta=
(\mathbf{d}h\wedge\zeta)\wedge\big[\mathbf{d}(ln\,f)+h\zeta\big]=
\big[\mathbf{d}(ln\,f)\big]\wedge\mathbf{d}h\wedge\zeta.
\]
Denoting for convenience $(ln\,f)=\varphi$ for the restriction $i^*\beta$ we
obtain
\[
i^*\beta=\varepsilon(\varphi_x h_y-\varphi_y h_x)dx\wedge dy\wedge dz.
\]

In order to find appropriate interpretation of $i^*\beta$ we easily check
that the 3-form
\begin{equation}
i(\bar{\zeta})(\mathbf{d}A\wedge A\wedge\zeta)
=i(\bar{\zeta})(\mathbf{d}A^*\wedge A^*\wedge\zeta)                 %15%
\end{equation}
looks like $\gamma\wedge\zeta$
with $\gamma=-\phi^2(\psi_\xi-\varepsilon\psi_z)dx\wedge dy$. Moreover, if
$$
\phi^2=\phi^2(x,y,\xi+\varepsilon z)\ \  \text{and} \ \
(\psi_\xi-\varepsilon\psi_z)=\pm\frac{1}{l_o}=const
$$
this 3-form is closed: $\mathbf{d}(\gamma\wedge\zeta)=0$.

Hence, the interpretation $i^*\beta=i^*(\gamma\wedge\zeta)$ requires
appropriate definition of the two functions $f$ and $h$, i.e. we must
have
\[
\varphi_x h_y -h_x\varphi_y=-\frac{\kappa}{l_o}\phi^2, \ \ \kappa=\pm 1.
\]
If we choose
\[
f=exp(\varphi)=exp\left[\int{\phi^2}dx\right],\ \
h=-\frac{\kappa}{l_o}y+const
\]
all requirements will be fulfilled, in particular,
$\varphi_{xy}=\varphi_{yx}=(\phi^2)_y$ and $h_{xy}=h_{yx}=0$.

The above choice of $f$ and $h$ allows the spatial restriction of the
Godbillon-Vey 3-form $\beta$ to be physically interpreted. So, the curvature
expressions and the rotational properties of the corresponding solutions can
be related to the integrability of the Pfaff system defined by $\zeta$, i.e.
to the straight-line translational propagation properties of the solutions.
Hence, a consistent translational-rotational propagation is possible.

\section{Lagrangian formulation}
First we consider a simpler but helpful and suggestive example of
a lagrangian defined by a complex valued function. Let the field of complex
numbers
$\mathbb{C}=(\mathbb{R}^2,J), J\circ J=-id_{\mathbb{R}^2}$ be given a real
representation as a 2-dimensional real vector space with basis
\[
I=\begin{Vmatrix}1 & 0 \\0 & 1\end{Vmatrix}, \ \ J=\begin{Vmatrix}0 & 1 \\-1
& 0\end{Vmatrix}.
\]
Note that the complex structure operator $J$ is a rotational operator
inducing rotation in $\mathbb{R}^2$ to the angle of $\pi/2$.
Every $\mathbb{C}$-valued function $\alpha$ on
$\mathbb{R}^4$ can be represented in the form
$\alpha=uI+pJ=\phi\cos\psi\,I+\phi\sin\psi\,J$,
where $u$ and $p$ are two
real-valued functions, $\phi=\sqrt{u^2+p^2}$ and $\psi=\arctan\frac pu$, and
the components of $\alpha$ with respect to this basis will be anumerated by
latin indices taking values $(1,2): \alpha_i, i=1,2$ . We denote further
$J(\alpha)=-pI+uJ\equiv \bar{\alpha}$. We introduce the two 1-forms (in our
$\zeta$-adapted coordinate system):  $k^s=i^*\zeta/l_o+0.d\xi$ and
$k^\xi=\xi^*\zeta/l_o+0.dx+0.dy+0.dz$ - the $l_o$-scaled $\xi$-restriction
of $\zeta$, with zero elements along the spatial directions, $l_o=const>0$.

Making use of our vector field $\bar{\zeta}$, of the inner product
$g$ in $(\mathbb{R}^2,J)$ defined by
$g(\alpha,\beta)=\frac12tr(\alpha\circ\beta^*)$
and working in the
corresponding $\bar\zeta$-adapted coordinate system $(x,y,z,\xi)$, we
consider now the following lagrangian (summation over the repeated indices:
$g(\alpha,\alpha)=\alpha_i\alpha_i$):
\[
\mathbb{L}=
\frac12\Big[\kappa l_o g(\alpha,L_{\bar{\zeta}}\bar{\alpha})+g(\alpha,\alpha)-
\kappa l_o
g(\bar{\alpha},L_{\bar{\zeta}}\alpha)+g(\bar{\alpha},\bar{\alpha})\Big]=
\]
\[
=\frac12\Big[\alpha_i\Big(\kappa l_o\bar{\zeta}^\sigma\frac{\partial
\bar{\alpha}_i}{\partial x^\sigma}+\alpha_i\Big)-
\bar{\alpha}_i\Big(\kappa
l_o\bar{\zeta}^\sigma\frac{\partial\alpha_i}{\partial
x^\sigma}-\bar{\alpha}_i\Big)\Big],
\]
where $l_o$ is of dimension [length], and $\kappa=\pm 1$. The Lagrange
equations are
\[
\kappa l_o
L_{\bar{\zeta}}\bar{\alpha}=-\alpha\ ;
\ \ \kappa l_o L_{\bar{\zeta}}\alpha=\bar{\alpha},
\]
or, in components,
\[
\kappa l_o\bar{\zeta}^\sigma\frac{\partial \bar{\alpha}_i}{\partial
x^\sigma}=-\alpha_i \ ;\ \
\kappa l_o\bar{\zeta}^\sigma\frac{\partial \alpha_i}{\partial
x^\sigma}=\bar{\alpha}_i.
\]
Note that the first equation follows from the second one under the
acting with $J$ from the left, hence, we have just one equation
of the form $\kappa l_o L_{\bar{\zeta}}\alpha=J(\alpha)$, which represents,
as we mentioned earlier, the idea for {\it consistent translational-rotational
propagation}. Moreover, since $J$ rotates to the angle of $\pi/2$, the
parameter $l_o$ fixes the corresponding translational advancement.

In terms of $(\phi,\psi)$ these equations give
\[
L_{\bar{\zeta}}\phi.\cos(\psi)
-\phi.\sin(\psi)\left(L_{\bar\zeta}\psi-
\frac{\varepsilon\kappa}{l_o}\right)=0, \ \
L_{\bar{\zeta}}\phi.\sin(\psi)
+\phi.\cos(\psi)\left(L_{\bar\zeta}\psi-
\frac{\varepsilon\kappa}{l_o}\right)=0,
\]
where
$\phi=\sqrt{\alpha_1^2+\alpha_2^2}$ and
$\psi=\arctan\frac{\alpha_2}{\alpha_1}$. These two equations are consistent
only if
\begin{equation}
L_{\bar{\zeta}}\phi=0,\ \
L_{\bar{\zeta}}\psi=\varepsilon\frac{\kappa}{l_o}.         %(16)
\end{equation}
So, equations (16) are our dynamical equations, and their solutions are:
\begin{equation}
\phi=\phi(x,y,\xi+\varepsilon z); \ \
\psi_1=
\psi_o(x,y,\xi+\varepsilon z)-\frac{\kappa}{l_o} z +C_1; \
\ \psi_2
=\psi_o(x,y,\xi+\varepsilon z)+
\frac{\kappa\varepsilon}{l_o}\xi+C_2,
\end{equation}
where $C_1$ and $C_2$ may depend on $(x,y)$.
Omitting the running wave term $\psi_o$ and assuming $C_1=const$ and
$C_2=const$ we see that
$$
\psi_1=-\frac{\kappa}{l_o} z+const=-k^s_\mu x^\mu+const \ \ \text{and} \ \
\psi_2=\frac{\kappa\varepsilon}{l_o} \xi+const =k^\xi_\mu x^\mu+const
$$
are the essential possible solutions leading to non-zero curvature
and, therefore, to available rotational component of propagation.

It is important to note that this lagrangian leads to {\it linear} equations
for the components of $\alpha$, which equations admit 3d-finite solutions
of the kind
$$
\alpha_1=\phi\cos\psi ; \ \
\alpha_2=\phi\sin\psi
$$
with consistent translational-rotational behavior, where
$\phi$ and $\psi$ are given above, and $\phi$ is a spatially finite function.
%Regarding the corresponding superposition of solutions we do not consider it
%as important and essential to be specially analysed because the various
%representations of the {\it same} $\mathbb{C}$-valued function, being
%solution or not-solution, through linear combinations of other
%$\mathbb{C}$-valued functions being also solutions or not-solutions,
%do NOT introduce anything new, in fact, the set of all such representations
%define the element under consideration. The set of solutions define just a
%subspace of all $\mathbb{C}$-valued functions, so, any solution can be
%represented in various ways as a linear combination of other solutions, and
%all such linear combinations define the same solution.

It is easily seen that the lagrangian becomes zero on the solutions, and
since this lagrangian does NOT depend on any space-time metric the
corresponding Hilbert energy-momentum tensor is zero on the solutions.  This
special feature of the lagrangian requires to look for another procedure
leading to corresponding conserved quantities. A good candidate seems to be
\linebreak
$T^{\mu\nu}=\phi^2 \bar{\zeta}^\mu\bar{\zeta}^\nu$. In fact, we obtain
\[
\nabla_\nu T^{\mu\nu}=\bar{\zeta}^\mu\nabla_\nu(\phi^2\bar{\zeta}^\nu)+
\phi^2 \bar{\zeta}^\nu\nabla_{\nu}\bar{\zeta}^\mu.
\]
The first term on the right is equal to $\bar\zeta^\mu
L_{\bar{\zeta}}\phi^2=0$, and the second term is zero because $\bar{\zeta}$
is autoparallel.

It is also easily seen that this lagrangian is $U(1)$-invariant. The
corresponding conserved quantity looks like $-\kappa l_oE$ where $E$ is the
integral energy, so, after deriving by the speed of light $c$ this quantity
becomes $\pm El_o/c = \pm ET$.

Most of these relations can be easily carried to a lagrangian defined in
terms of a 2-form in the following way. Recall the complex structure
$\mathcal{J}$ and define a representation $\rho$ of the algebra
$\mathbb{C}$ in the algebra $L_{\Lambda^2(\mathbb{R}^4)}$ of linear
maps in the 2-forms on $\mathbb{R}^4$ by the relation
\begin{equation}
\rho(\alpha_\varepsilon)=\rho(uI+\varepsilon pJ)\overset{\text{\small def}}{=}
u\mathcal{I}+\varepsilon p\mathcal{J}, \ \                     %18%
\mathcal{I}=id_{\Lambda^2(\mathbb{R}^4)},
\ \ \alpha_{\varepsilon}\in\mathbb{C}, \ \ \varepsilon=\pm 1.
\end{equation}
Clearly,
$\rho(\alpha+\beta)=\rho(\alpha)+\rho(\beta), \  \rho(\alpha.\beta)=
\rho(\alpha)\circ\rho(\beta)$, and if $F$ is an arbitrary 2-form
$\rho(\alpha_\varepsilon).F=uF+\varepsilon p\mathcal{J}(F)$.
Note that here and further in the text the
couple $(u,p)$ may denote the complex number $(uI+pJ)$, as well as the
complex-valued function $\alpha=u(x,y,z,\xi)I+p(x,y,z,\xi)J$.

Let's go back now to our $\bar\zeta$-adapted coordinate system and consider
the 2-form $F_o=dx\wedge\zeta=\varepsilon dx\wedge dz+dx\wedge\xi$. We obtain
\[
\rho(\alpha_{\varepsilon}).F_o=\varepsilon udx\wedge dz+
\varepsilon pdy\wedge dz+udx\wedge d\xi+pdy\wedge d\xi=A\wedge\zeta,
\]
\[
\rho(J(\alpha_{\varepsilon})).F_o=
(-\varepsilon p\mathcal{I}+u\mathcal{J}).F_o=
\]
\[
=-pdx\wedge dz+
udy\wedge dz-\varepsilon pdx\wedge d\xi+\varepsilon udy\wedge d\xi
=A^*\wedge\zeta=
\mathcal{J}(\rho(\alpha_\varepsilon).F_o).
\]
In this way we get one-to-one map between the $\mathbb{C}$-valued functions
on $\mathbb{R}^4$ and a special subset of 2-forms. All such 2-forms depend
on the choice of the 1-form $\zeta$, while the dependence on $dx$ is not
essential.  Also, they are isotropic, i.e. they have zero invariants.

Moreover, every such 2-form may be considered as a linear map in
$\Lambda^2(\mathbb{R}^4)$ through the above correspondence:
$\rho(\alpha_\varepsilon).F_o\rightarrow \rho(\alpha_\varepsilon)$.
Since together with the zero element of $\Lambda^2(M)$ these 2-forms define a
linear space $V_\zeta$, this property suggests to introduce inner product in
this linear space by the rule $<F^1_\varepsilon (a,b), F^2_\varepsilon
(m,n)>= \frac16 tr\Big[\rho\big[\alpha_\varepsilon(m,n)\big]\circ
\rho\big[\alpha^*_\varepsilon (a,b)\big]\Big]=am+bn$. Hence, every such
2-form acquires a norm. Note that this inner product coincides with the
induced by a pseudoeuclidean metric $\eta$ inner product $g$ on the homology
space generated by the corresponding energy-tensor as we mentioned in the
preceding section. But, the richer nature of the inner product $<,>$
is manifested in that it generates a Hermitian one in $V_\zeta$ with respect
to the natural product
$$
\lambda_\varepsilon.F_\varepsilon(u,p)=
\rho(\lambda_\varepsilon)\circ\rho\big[\alpha_\varepsilon(u,p)\big]=
\rho\big[\lambda_\varepsilon.\alpha_\varepsilon(u,p)\big], \ \
\lambda_\varepsilon=const\in\mathbb{C},
$$
under the correspondence
$F_\varepsilon(u,p)\rightarrow\rho(\alpha_\varepsilon(u,p))$.

Let now $F$ and $G$ be two arbitrary 2-forms. In order to define the
lagrangian we consider the Minkowski space-time $M=(\mathbb{R}^4,\eta)$ as a
real manifold, where the pseudoeuclidean metric $\eta$ has signature
$(-,-,-,+)$, and make use of the Lie derivative $L_{\bar\zeta}$ with respect
to the vector field $\bar\zeta$. Also, $-(k^s)^2=(k^\xi)^2=(l_o)^{-2}$. The
lagrangian is given by
\[
\mathbb{L}=\eta\left(\kappa l_oL_{\bar{\zeta}}G+F,F\right)-
\eta\left(\kappa l_oL_{\bar{\zeta}}F-G,G\right)=
\]
\begin{equation}
=\frac12\left(\kappa l_o\bar{\zeta}^\sigma
\frac{\partial G_{\alpha\beta}}{\partial x^\sigma}+
F_{\alpha\beta}\right)F^{\alpha\beta}-
\frac12\left(\kappa l_o\bar{\zeta}^\sigma                          %19%
\frac{\partial F_{\alpha\beta}}{\partial x^\sigma}-
G_{\alpha\beta}\right)G^{\alpha\beta}, \ \ 0<l_o=const, \ \ \kappa=\pm 1,
\end{equation}
where $F^{\alpha\beta}=\eta^{\alpha\mu}\eta^{\beta\nu}F_{\mu\nu}$ and
$G^{\alpha\beta}=\eta^{\alpha\mu}\eta^{\beta\nu}G_{\mu\nu}$. Note that this
lagrangian is invariant with respect to $(F,G)\rightarrow(G,-F)$, or to
$(F,G)\rightarrow(-G,F)$.

The corresponding equations read
\[
\kappa l_o\bar{\zeta}^\sigma
\frac{\partial G_{\alpha\beta}}{\partial x^\sigma}+
F_{\alpha\beta}=0, \ \
\kappa l_o\bar{\zeta}^\sigma
\frac{\partial F_{\alpha\beta}}{\partial x^\sigma}-
G_{\alpha\beta}=0.
\]
In coordinate-free form these equations look like
\[
\kappa l_oL_{\bar{\zeta}}G=-F, \ \ \ \kappa l_oL_{\bar{\zeta}}F=G.
\]

Recall now the complex structure $\mathcal{J}$ and assume $G=\mathcal{J}(F)$.
Then, treating $F$ and $\mathcal{J}(F)$ as independent (in fact they are
linearly independent on the real manifold $M$) we can write
\begin{equation}
\kappa l_oL_{\bar{\zeta}}\mathcal{J}(F)=-F, \ \ \ \kappa
l_oL_{\bar{\zeta}}F=\mathcal{J}(F) .                          %(20)%
\end{equation}
Since in our coordinates $\mathcal{J}$ and $\bar\zeta$ have constant
coefficients, clearly, $L_{\bar\zeta}$ and $\mathcal{J}$ commute, so that the
second (first) equation is obtained by acting with $\mathcal{J}$ from the
left on the first (second) equation, i.e. the above mentioned invariance with
respect to the transformations $(F,G)\rightarrow(\pm G,\mp F)$ is reduced to
$\mathcal{J}$-invariance. Recalling now the way $\mathcal{J}$ acts,
$\mathcal{J}(A\wedge\zeta)=A^*\wedge\zeta$, i.e. the couple $(A,\zeta)$ is
rotated to the couple $(A^*,\zeta)$, we naturally interpret the last
equations as realization of the {\it translational-rotational consistency}:
the translational change of $F$ along $\bar\zeta$ is proportional to the
rotational change of $F$ determined by $\mathcal{J}$, so, roughly speaking,
{\bf no $\bar\zeta$-translation ($\mathcal{J}$-rotation) is possible without
$\mathcal{J}$-rotation ($\bar\zeta$-translation)}, and the
$\mathcal{J}$-rotation corresponds to $l_o$ translational advancement.

From (20) it follows that on the solutions the lagrangian becomes zero.
So, if we try to define the corresponding Hilbert energy-momentum tensor
the variation of the volume element with respect to $\eta$ is not essential.
Moreover, the special quadratic dependence of $\mathbb{L}$ on $\eta$ shows
that the variation of $\mathbb{L}$ with respect to $\eta$ will also become
zero on the solutions. Hence, this is another example of the non-universality
of the Hilbert method to define appropriate energy-momentum tensor. As for
the canonical energy-momentum tensor, it is not symmetric, and its
symmetrization is, in some extent, an arbitrary act, therefore, we shall
not make use of it.

In order to come to more realistic relations we shall restrict equations (20)
on the subset of 2-forms $F$ of the above defined kind, namely,
$F=\rho(\alpha_\varepsilon).F_o$. All these 2-forms have zero invariants:
$F_{\mu\nu}F^{\mu\nu}=F_{\mu\nu}(\mathcal{J}(F))^{\mu\nu}=0$, or in
coordinate-free way, $F\wedge F=F\wedge\mathcal{J}(F)=0$. Moreover, the
easily verified relations $i(\bar\zeta)F=i(\bar\zeta)\mathcal{J}(F)=0$ show
an {\it intrinsic} connection to $\bar\zeta$: it is the only
isotropic eigen vector of $F_\mu^\nu=\eta^{\mu\sigma}F_{\nu\sigma}$ and
$(\mathcal{J}F)_\mu^\nu=\eta^{\mu\sigma}(\mathcal{J}F)_{\nu\sigma}$.

Substituting $F=\rho(\alpha_\varepsilon).F_o$ in (20) we get the equations
\begin{equation}
\kappa l_o (u_\xi-\varepsilon u_z)=-\varepsilon p, \ \ \          %21%
\kappa l_o(p_\xi-\varepsilon p_z)=\varepsilon u,
\end{equation}
i.e. $\kappa l_oL_{\bar\zeta}u=-\varepsilon p$ and
$\kappa l_oL_{\bar\zeta}p=\varepsilon u$.  From these equations
we obtain the relations
\begin{equation}
(u^2+p^2)_\xi-\varepsilon                                        %22%
(u^2+p^2)_z=\frac12L_{\bar\zeta}(u^2+p^2)=0, \ \
u(p_\xi-\varepsilon p_z)-p(u_\xi-\varepsilon u_z)
=\frac{\varepsilon\kappa}{l_o}(u^2+p^2).
\end{equation}
The substitution $u=\phi\,\cos(\psi), \ \ p=\phi\,\sin(\psi)$ in (22) leads
to the equations (16)
\[
L_{\bar\zeta}\phi=0, \ \ \ L_{\bar\zeta}\psi=\frac{\varepsilon\kappa}{l_o},
\]
and to the corresponding solutions (17). Now, relations (11),(13),(14) and
(22) suggest to choose $\phi^2\zeta\otimes\bar\zeta$ for energy-momentum
tensor.

Note that the 2-form $F_o=dx\wedge\zeta$ satisfies the first of
these equations since $\phi_{F_{o}}=1$ and does NOT satisfy the second
equation since $\psi_{F_o}=0,2\pi,4\pi, ...$, so,
$L_{\bar\zeta}(\psi_{F_o})=0$. In view of this further we consider only
not-constant $\mathbb{C}$-valued functions.

As for integral characteristic of the intrinsic rotational properties of a
solution a good candidate seems to be the integral over $\mathbb{R}^3$ of the
restricted to $\mathbb{R}^3$ appropriately interpreted representative
of the Godbillon-Vey class as given by relation (15). Making use of the
solution $\psi_1$ as given above and choosing appropriate coefficient, we
obtain
\[
\beta=\frac{2\pi (l_o)^2}{c}
\Big[-\varepsilon\phi^2(\psi_\xi-\varepsilon\psi_z)dx\wedge dy\wedge dz-
\phi^2(\psi_\xi-\varepsilon\psi_z)dx\wedge dy\wedge d\xi\Big],
\]
so,
\[
i^*\beta=2\pi\frac{l_o}{c}\kappa\phi^2 dx\wedge dy\wedge dz.
\]
Since the energy-density is given by the spatially finite function $\phi^2$
the corresponding integral of $i^*\beta$ over $\mathbb{R}^3$ gives
$\kappa ET=\pm ET$, where $E$ is the integral energy of the solution and
$T=2\pi l_o/c$.

Another also appropriate local representative of the rotational properties
of these solutions appears to be any of the two Frobenius 4-forms
$\mathbf{d}A\wedge A\wedge A^*$ and  $\mathbf{d}A^*\wedge A^*\wedge A$,
multiplied by the coefficient $2\pi\varepsilon l_o/c$, so that integrating
over the 4-region $(\mathbb{R}^3\times l_o)$ we get also $\kappa ET$.

The linear character of the equations obtained sets the question if the
superposition principle holds. In general, let the parameters
$\kappa, \varepsilon, l_o$ of the two solutions be different. Let now
$F_1(\kappa_1,\varepsilon_1,l_o^1;u,p)$ and
$F_2(\kappa_2,\varepsilon_2,l_o^2;m,n)$ be two solutions along the same
direction defined by $\bar\zeta$, and $\varepsilon$ of $\zeta$ is of
course equal to $\varepsilon_1$ for the first solution, and equal to
$\varepsilon_2$ for the second solution.  We ask now whether the linear
combination $c_1F_1+c_2F_2$ with $c_1=const, c_2=const$ will be also a
solution along the same direction?  In order this to happen the following
equations must be consistent:
\[
\kappa_1\varepsilon_1l_o^1L_{\bar\zeta}u=-\varepsilon_1 p, \ \
\kappa_1\varepsilon_1l_o^1L_{\bar\zeta}p=\varepsilon_1 u, \ \
\kappa_2l_o^2L_{\bar\zeta}m=-\varepsilon_2 n, \ \
\kappa_2l_o^2L_{\bar\zeta}n=\varepsilon_2 m
\]
\[
\kappa_3\varepsilon_3l_o^3L_{\bar\zeta}(c_1u+c_2m)=-\varepsilon_3(c_1p+c_2n),
\ \ \kappa_3\varepsilon_3l_o^3L_{\bar\zeta}(c_1p+c_2n)=
\varepsilon_3 (c_1u+c_2m),
\]
where $\varepsilon_3$ is equal $\varepsilon_1$, or to $\varepsilon_2$.
The
corresponding consistency condition looks as follows:
\[
\kappa_3\varepsilon_3 l_o^3
=\frac{c_1p+c_2n} {\frac{\varepsilon_1\kappa_1c_1}{l_o^1}p+
\frac{\varepsilon_2\kappa_2c_2}{l_o^2}n}=
\frac{c_1u+c_2m}
{\frac{\varepsilon_1\kappa_1c_1}{l_o^1}u+
\frac{\varepsilon_2\kappa_2c_2}{l_o^2}m}.
\]
For example, the relations $\varepsilon_3\kappa_3l_o^3=
\varepsilon_2\kappa_2l_o^2=\varepsilon_1\kappa_1l_o^1$ are sufficient for
this superposition to be a solution. This means that if the two solutions
propagate translationally for example from $-\infty$ to $+\infty$, i.e.
$\varepsilon_1=\varepsilon_2=-1$, if the rotational orientations coincide,
i.e. $\kappa_1=\kappa_2$, and if the spatial periodicity parameters are equal:
$l_o^1=l_o^2=l_o^3$, the sum $(c_1F_1+c_2F_2)$ gives a solution.  In general,
however, the combination $c_1F_1+c_2F_2$ will not be a solution.

On the other hand, we can introduce a multiplicative structure in the
solutions of the kind $\rho(\alpha_\varepsilon).F_o$. In fact, if
$F_1(\kappa_1)=\rho(\alpha^1(\varepsilon_1)).F_o$ and
$F_2(\kappa_2)=\rho(\alpha^2(\varepsilon_2)).F_o$ are two such
solutions we define their product $F=F_1.F_2$ by
\[
F=F_1.F_2=\rho(\alpha^1.\alpha^2).F_o.
\]
Clearly, the amplitude of $F$ is a product of
the amplitudes of $F_1$ and $F_2$: $\phi_F=\phi_{F_1}.\phi_{F_2}$ and the
phase of $F$ is the sum of the phases of $F_1$ and $F_2$:
$\psi_F=\psi_{F_1}+\psi_{F_2}$. Now, $F$ will be a solution only if
$$
\varepsilon_1=\varepsilon_2=\varepsilon_{F}, \ \
\frac{\kappa_{F}}{l_o^{F}}=\frac{\kappa_1l_o^2+\kappa_2l_o^1}{l_o^1.l_o^2}.
$$
From the first of these relations it follows that in order $F$ to be a
solution, $F_1$ and $F_2$ must NOT move against each other, and then the
product-solution shall follow the same translational direction. However, it
is allowed $F_1$ and $F_2$ to have different rotational orientations, i.e.
$\kappa_1\neq\kappa_2$, then the product-solution will have rotational
orientation $\kappa_F=sign(\kappa_1l_o^2+\kappa_2l_o^1)$. Clearly, every
subset of solutions with the same $(\varepsilon,\kappa,l_o)$
 form a group with neutral element $F_o=\rho(I).F_o$, and
$F_\alpha^{-1}=\rho(\alpha^{-1}).F_o$,  moreover, the multiplicative group of
$\mathbb{C}$ acts on these solutions:  $(\alpha_\varepsilon,
F_\varepsilon)\rightarrow \rho(\alpha_\varepsilon).F_\varepsilon$.

\section{Discussion and Conclusion}
The main feature of individual objects of classical mechanics is that they
admit the approximation {\it material point}, i.e. the physical situations
considered in the frame of classical mechanics do NOT admit these objects to
be created and destroyed, they just change their state of motion under
external influences, so, as theoretical entities, they are {\it uncreatable}
and {\it indestructible}.  In other words, their structure is neglected, they
are considered as structureless.

On the contrary, photons appeared in physics as {\it creatable} and {\it
destructible} objects, i.e. objects with structure, hence, a new kind of
theory describing their behavior in accordance with their structure was
needed.  Although Einstein's interpretation of Planck's formula concerning
light quanta in view of the relativity principle, the later developed quantum
mechanics continued to consider the point-like approximation of microobjects
as admissible and consistent in the context of some probabilistic
interpretation.  Moreover, a general understanding of the physical situation
on the base of some definite {\it uncertainty/complementarity relations}
was formulated.  Quantum electrodynamics also has not presented so far a
realistic spatially finite and time-stable model of individual photons with
dynamical structure.

The two basic features of our approach in this paper are the assumptions for
continuous spatially finite structure, and for available consistent
translational-rotational dynamical structure of photons. Hence, photons {\it
propagate}, they do not move. The propagation has two components: {\it
translational} and {\it rotational}. The translational component is along
isotropic straight lines in $(\mathbb{R}^4,\eta)$.  The rotational component
of propagation follows the special rotational properties of photon's spatial
structure.  This dual nature of photons demonstrates itself according to the
rule: {\bf no translation (rotation) is possible without
rotation (translation)} (equation (20)).

While the translational component of propagation is easily accounted through
the (arbitrary chosen in general) isotropic autoparallel vector field
$\bar\zeta$, the rotational component of propagation was introduced in
our model making use of two things: the non-integrability properties of the
induced by $\bar\zeta$ two 2-dimensional differential/Pfaff systems, and the
complex structure $\mathcal{J}$.  This approach brought the following
important consequences:

1. It automatically led to the required translational-rotational
consistency.

2. The rotational properties of the solutions have {\it invariant} and
{\it intrinsic} for the object nature, they are transversal to $\bar\zeta$,
they are in accordance with the action of the complex structure $\mathcal{J}$
on the 2-forms $G$ and $G^*$, and they allow the
characteristics {\it amplitude} $\phi$ and {\it phase} $\psi$ of a solution
to be correctly introduced.  Moreover, the rotation is of periodical nature,
and a helical spatial structure along $\bar\zeta$ is allowed, so, the
corresponding rotational properties differ from those in the case of rotation
of a solid as a whole around a point or axis.

3. The translational properties of a solution are carried by the amplitude
$\phi :\ L_{\bar\zeta}\phi=0$, and the rotational ones - by the phase
$\psi$: $L_{\bar\zeta}\psi\neq 0$.

4. The curvature, considered as a measure of the available non-integrability,
is non-zero only if the phase $\psi $ is NOT a running wave along
$\bar\zeta:$\ $L_{\bar\zeta}\psi\neq 0 $, hence, curvature means rotation and
vice versa.

5. Quantitatively, the curvature is {\it obtained} to be proportional to the
energy-density $\phi^2$ (and to the change of the phase $\psi$ along
$\bar\zeta$) of a solution.  We recall that in General Relativity a
proportionality relation between the energy-density of non-gravitational
fields and the corresponding (contracted) riemannian curvature is {\it
postulated}, while here it is {\it obtained} in the most general sense of the
concept of curvature, namely, as a measure of Frobenius non-integrability.

6. A natural {\it integral} measure of the rotational properties of a
solution appears to be the product $ET$, i.e. the action for one period
$T=2\pi l_o/c$, which is in accordance with the Planck formula $ET=h$. It
seems remarkable that this quantity may be obtained also through the spatial
restriction of the available and correspondingly interpreted Godbillon-Vey
class.

Together with the allowed finite nature of the solutions these properties
suggest the following understanding of the photons' time-stable dynamical
structure:  {\bf photons MUST always propagate in a translational-rotational
manner as fast as needed in order to "survive", i.e.  to overcome the
instability (the destroying tendencies), generated by the available
non-integrability}. In other words, every free photon has to be able to
supply immediately itself with those existence needs that are constantly
put under the non-integrability destroying influence. Some initial steps to
understand quantitatively this "smart" nature of photons in the terms used in
the paper could be the following.

Recalling the two 2-forms $G=A\wedge\zeta$ and
$G^*=A^*\wedge\zeta$ and relations (11)-(12) we see that when
$L_{\bar\zeta}\phi^2=0$ then $G$ and $G^*$ keep the
energy-momentum carried by each of them:
$i(\bar{G})\mathbf{d}G=i(\bar{G}^*)\mathbf{d}G^*=0$.
On the other hand the relation
$i(\bar{G})\mathbf{d}G^*=-i(\bar{G}^*)\mathbf{d}G=\mathbf{R}.\zeta $
may be physically interpreted in two ways. FIRST, differentially, $G$
transfers to $G^*$ so much energy-momentum as $G^*$ transfers back to $G$,
which goes along with the previous relations stating that $G$ and $G^*$ keep
their energy-momentum densities. Each of these two quantities
$i(\bar{G})\mathbf{d}G^*$ and $i(\bar{G}^*)\mathbf{d}G$
is equal (up to a sign) to $\mathbf{R}.\zeta$, so, such mutual exchange of
energy-momentum is possible only if the non-integrability of each of the two
Pfaff 2-dimensional systems $(A,\zeta)$ and $(A^*,\zeta)$ is present and
measured by the same non-zero quantity, i.e. when the curvature $\mathbf{R}$
is NOT zero. Since the curvature implies outside/inside directed flow, this
suggests the SECOND interpretation:  the energy-momentum that photons lose
differentially in {\it whatever way} by means of $G$ is differentially
supplied by means of $G^*$, and vice versa.  We could say that every photon
has two functioning subsystems, $G$ and $G^*$, such, that the
energy-momentum loss through the subsystem $G$ generated by the
nonintegrability of $(A,\zeta)$, is gained (or supplied) back by the subsystem
$G^*$, and vice versa, and in doing this photons make use of the
corresponding rotational component of propagation supported by appropriate
spatial structure. All this is mathematically guaranteed by the isotropic
character of $G$ and $G^*$, i.e.  by the zero values of the two invariants
$G_{\mu\nu}G^{\mu\nu}=G^*_{\mu\nu}G^{\mu\nu}=0$, and by making use of the
complex structure $\mathcal{J}$ as a rotation generating operator.

Let's now try to express this dually consistent dynamical nature of photons
by {\it one} object which satisfies {\it one} relation. We are going to
consider $G$ and $G^*$ as two vector components of a
$(\mathbb{R}^2,J)$-valued 2-form, namely, $\Omega=G\otimes I+G^*\otimes J$.
Applying the exterior derivative we get $\mathbf{d}\Omega=\mathbf{d}G\otimes
I+\mathbf{d}G^*\otimes J$. Consider now the $(\mathbb{R}^2,J)$-valued
2-vector $\bar\Omega=\bar{G}\otimes I+\bar{G^*}\otimes J$. The aim we pursue
will be achieved through defining the object ($\vee$ is the symmetrized
tensor product)
\[ (\vee,i)(\bar\Omega,\mathbf{d}\Omega)\overset{\text{\small
def}}{=} i(\bar{G})\mathbf{d}G\otimes I\vee I +
i(\bar{G^*})\mathbf{d}G^*\otimes J\vee J+
\Big[i(\bar{G})\mathbf{d}G^*+i(\bar{G^*})\mathbf{d}G\Big]\otimes I\vee J
\]
and put it equal to zero: $(\vee,i)(\bar\Omega,\mathbf{d}\Omega)=0$.

Finally we note that this last relation
$(\vee,i)(\bar\Omega,\mathbf{d}\Omega)=0$ represents the dynamical
equations of the vacuum Extended Electrodynamics [24]. In particular,
this equation contains all solutions to the Maxwell vacuum equations, and
the solutions obtained in this paper are a special part of the full subset of
nonlinear solutions to these nonlinear equations.

In conclusion, spatially finite field models of photons can be built in terms
of complex valued functions and in terms of 2-forms on Minkowski space-time,
and these two approaches can be related.  In these both cases substantial
role play the complex structures $J$ and $\mathcal{J}$ as rotation generating
operators, carrying information in this way about the spin properties of
photons. This moment throws some light on the long standing and nonanswered
question why quantum theory works with $(\mathbb{R}^2,J)$-valued objects and
not just with $\mathbb{R}^2$-valued ones:  may be just the spin properties of
micro-objects, induced by their dynamical structure, are, at least partially,
accounted for in this way.

The photons' longitudinal dimensions and rotational orientations can be
determined by the constant parameter combination $\kappa l_o$, and their
transversal dimensions depend on the energy-density. The time-stability is
guaranteed, on one hand, by the internal energy-momentum exchange between the
two non-integrable 2-dimensional differential/Pfaff systems, and on the other
hand, by a dynamical harmony with the outside world. The models obtained are
consistent with the present day knowledge about the propagational behavior
and the integral energy-momentum and spin characteristics of photons.

We kindly acknowledge the support of the Bulgarian Science Research Fund
through Contract $\Phi/15/15$.

\newpage
%\vskip 1.5cm
{\bf References}
\vskip 0.5cm
[1]. {\bf M. Planck}, Ann. d. Phys., {\bf 4}, 553 (1901)

[2]. {\bf A. Einstein}, Ann. d. Phys., {\bf 17}, 132 (1905)

[3]. {\bf L. V. De Broglie}, "Ondes et quanta", C. R. {\bf 177}, 507 (1923)

[4]. {\bf E. Schr\"odinger}, Ann. d. Phys., {\bf 79}, 361 (1926); {\bf 79},
489 (1926); {\bf 80}, 437 (1926); {\bf 81}, 109 (1926)

[5]. {\bf G. N. Lewis}, Nature, {\bf 118}, 874 (1926)

[6]. {\bf P. Speziali}, Ed. {\it Albert Einstein-Michele Besso
Correspondence} (1903-1955), Herman, Paris pg.453 (1972)

[7]. {\bf M.Planck}, J.Franklin Institute, 1927 (July), p.13.

[8]. {\bf J.J.Thomson}, Philos.Mag.Ser. 6, {\bf 48}, 737 (1924), {\bf 50},
1181 (1925), and Nature, {\bf 137}, 23 (1936).

[9]. {\bf N.Rashevsky}, Philos.Mag. Ser.7, {\bf 4}, 459 (1927).

[10]. {\bf B. Lehnert, S. Roy}, {\it Extended Electromagnetic Theory}, World
Scientific, 1998.

[11]. {\bf G.Hunter,R.Wadlinger}, Phys.Essays, {\bf 2}, 158 (1989).

[12]. {\bf W. Honig}, Found. Phys., {\bf 4}, 367 (1974).

[13]. {\bf A.Lees}, Phyl.Mag., {\bf 28}, 385 (1939).

[14]. {\bf N.Rosen}, Phys.Rev., {\bf 55}, 94 (1939).

[15]. {\bf D.Finkelstein, C.Misner}, Ann.Phys., {\bf 6}, 230 (1959).

[16]. {\bf D.Finkelstein}, Journ.Math.Phys., {\bf 7}, 1218 (1966).

[17]. {\bf J.P.Vigier}, Found. of Physics, {\bf 21}, 1991 (125).

[18]. {\bf A. C. de la Torre}, arXiv:quant-ph/0410179

[19]. {\bf V. Krasnoholovets}, arXiv: quant-ph/0202170

[20]. {\bf J. Dainton}, Phil. Trans. R. Soc. Lond., {\bf A 359} (2000), 279

[21] {\bf H. Stumpf, T. Borne}, Annales de la Fond. Louis De Broglie, {\bf
26}, No. {\it special}, 2001 (429)

[22]. {\bf R. M. Godbole}, arXiv: hep-th/0311188

[23]. {\bf R. Nisius}, arXiv: hep-ex/0110078

[24] {\bf Donev, S., Tashkova, M.}, Proc. R. Soc. Lond., {\bf A 450}, 281
(1995); hep-th/0403244

\end{document}